\documentclass{rnaastex}
\usepackage{enumitem}
\usepackage{hyperref}

\setlist{nolistsep}

\begin{document}

\title{Python Leap Second Management and Implementation of Precise Barycentric Correction (barycorrpy)}

\correspondingauthor{Shubham Kanodia}
\email{szk381@psu.edu}

\author{Shubham Kanodia}
\affiliation{Department of Astronomy \& Astrophysics and Center for Exoplanets and Habitable Worlds, 525 Davey 
Laboratory, The Pennsylvania State University, University Park, PA, 16802, USA}
\author{Jason Wright}
\affiliation{Department of Astronomy \& Astrophysics and Center for Exoplanets and Habitable Worlds, 525 Davey 
Laboratory, The Pennsylvania State University, University Park, PA, 16802, USA}



\vspace{-1.5cm}\section{Precise Barycentric Corrections in Python} 
We announce {\tt barycorrpy} (BCPy)\footnote{\url{https://github.com/shbhuk/barycorrpy}}, a Python implementation to calculate precise barycentric corrections well below the 1 cm/s level, following the algorithm of \citet{JandJ}. This level of precision is required in the search for 1 M$_\Earth$ planets in the Habitable Zones of Sun-like stars by the Radial Velocity (RV) method, where the maximum semi-amplitude is $\approx$ 9 cm/s. We have developed BCPy to be used in the pipeline for the next generation Doppler Spectrometers - Habitable-zone Planet Finder (HPF \citet{hpf}) and NEID (\citet{neid,neid2}).

 BCPy is adapted from {\tt barycorr.pro}, the IDL implementation of the \citet{JandJ} algorithm. Prior to this there did not exist a public Python version of {\tt barycorr.pro} (there is a Python applet to query the {\tt barycorr.pro} web applet\footnote{\url{http://astroutils.astronomy.ohio-state.edu/exofast/barycorr.html}}, however not only is this process slow, but also requires Internet access. BCPy is about 5x faster than querying the applet for the same input parameters).  
 
 Astropy\footnote{\url{https://github.com/astropy/astropy/pull/6861}} has a barycentric correction routine which was recently updated to include some factors like gravitational time dilation, and Earth precession-nutation. This can reach 1 cm/s precision, however it does not include the effects of stellar proper motion or systemic radial velocity. Hence we propose BCPy as an effective tool to obtain barycentric velocity corrections precise below the 1 cm/s level. 

For the barycentric correction in BCPy we include the following effects : 
\begin{itemize}
\item Revolution of the Earth to consider position and velocity of the geocenter with respect to the Solar System barycenter
\item Rotation of the Earth
\item Precession, nutation and polar motion of the Earth, along with the above to calculate the position and velocity of the observatory with respect to the geocenter
\item Gravitational time dilation due to objects of the Solar System
\item Leap second offset
\item Proper motion and systemic radial velocity of the star
\item Parallax
\item Shapiro delay
\end{itemize}

\begin{figure}
\href{https://github.com/shbhuk/barycorrpy/blob/master/barycorrpy/tempo_ephem_checkv5.txt}
  \centering
  \begin{tabular}[b]{@{}p{0.40\textwidth}@{}}
  \href{https://github.com/shbhuk/barycorrpy/blob/master/barycorrpy/tempo_ephem_checkv5.txt}{
    \centering\includegraphics[width=\linewidth]{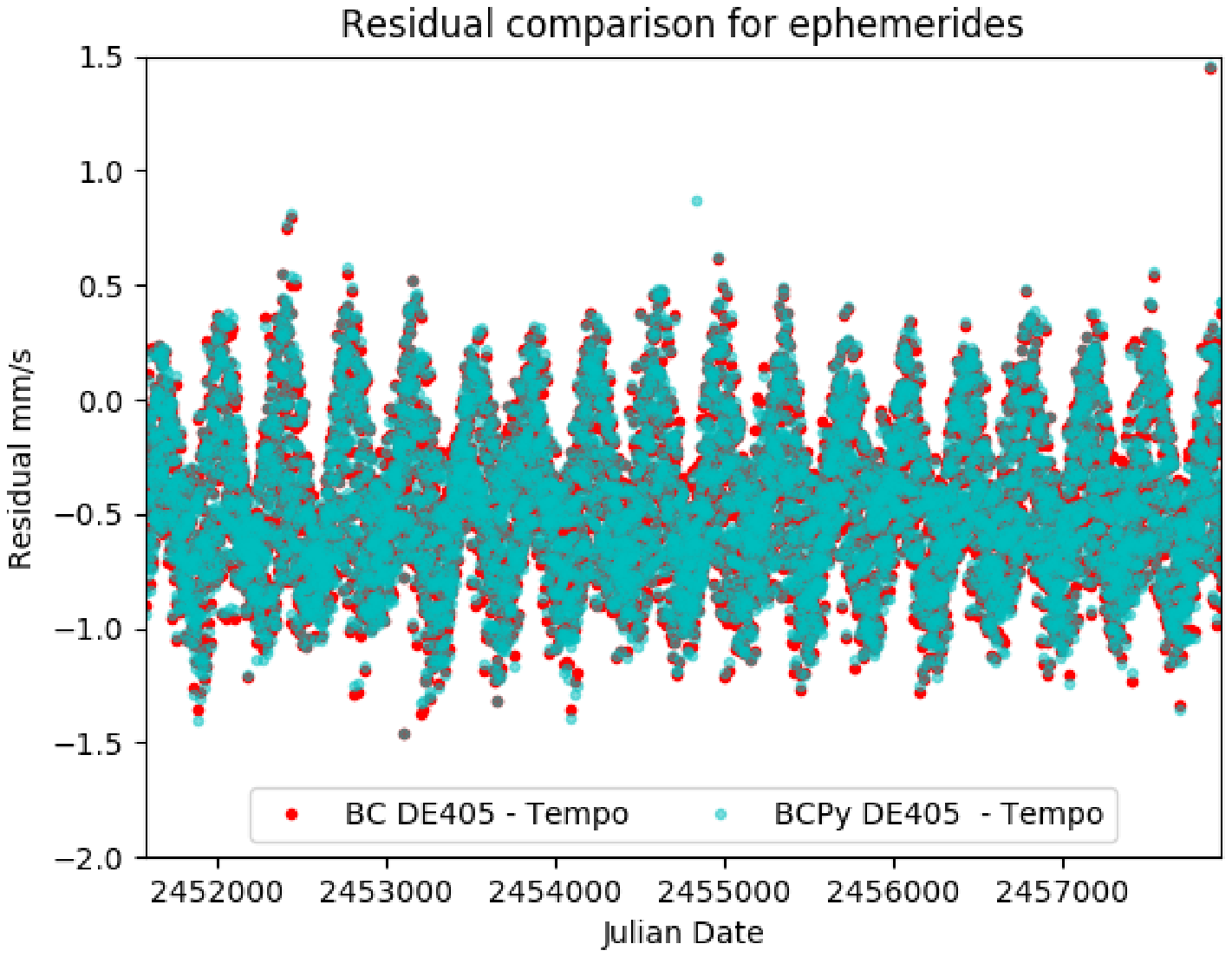}} \\
    \centering\small (a) Plot showing the residuals on comparing with {\tt TEMPO2}
  \end{tabular}%
  \quad
  \begin{tabular}[b]{@{}p{0.40\textwidth}@{}}
  \href{https://github.com/shbhuk/barycorrpy/blob/master/barycorrpy/tempo_ephem_checkv5.txt}{
    \centering\includegraphics[width=\linewidth]{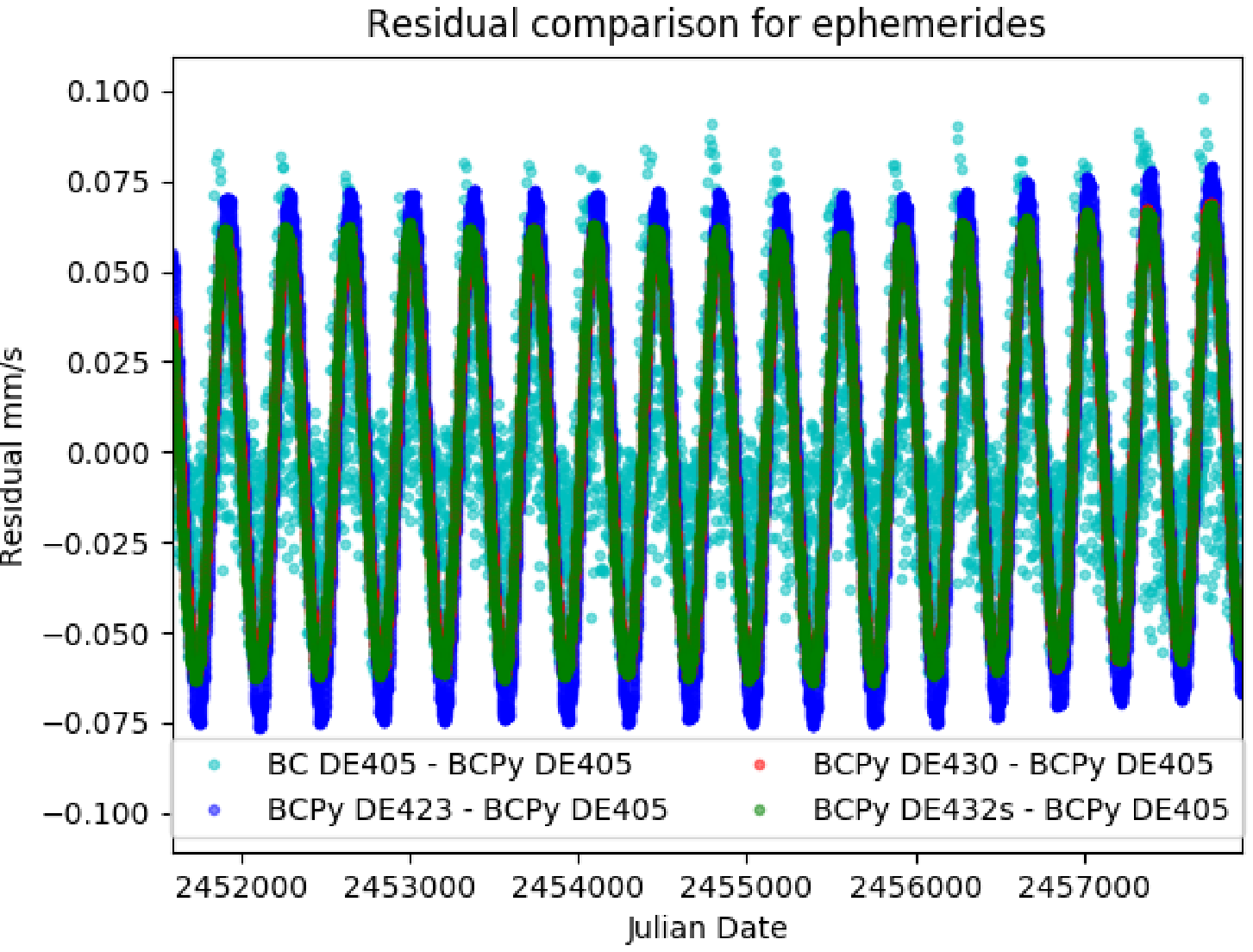}} \\
    \centering\small (b) Plot showing the residuals on comparing with \texttt{Barycorr.pro}
  \end{tabular}
  \caption{The various ephemerides available with BCPy compared to {\tt TEMPO2}  and with \texttt{barycorr.pro}. Here BC refers to \texttt{barycorr.pro}. Both BCPy and {\tt barycorr.pro} agree with {\tt TEMPO2} to better than 1 mm/s, and with each other to better than 0.1 mm/s, which is comparable to the uncertainties in the ephemerides.}   \label{fig:compare}
\end{figure}

In Figure \ref{fig:compare}
which is analogous to Table 2 from \cite{JandJ}, we compare the results of BCPy to {\tt TEMPO2} (\citet{Tempo}) and {\tt barycorr.pro} using the same input parameters for $\tau$ Ceti observed from the Cerro Tololo Inter-American Observatory (CTIO). Further, we also include a comparison for the various JPL ephemerides which we make available as options in this code. The data behind the figure is available on our GitHub repository. 

On our GitHub repository wiki we discuss the precision in the input parameters required to obtain 1 cm/s precision in the barycentric correction velocity. An important parameter is the time of observation, which is usually in JD UTC format. However, when querying the ephemerides to find the position of the Solar system bodies we need to convert the UTC time scale to TDB (Barycentric Dynamic Time) which requires precise accounting of leap seconds.


\section{Leap Second Management in Python}

In this work, we improve upon the Leap second management in Astropy. Since the Essential Routines for Fundamental Astronomy (ERFA) library that Astropy uses for this purpose has deliberately been kept similar to the Standards of Fundamental Astronomy (SOFA) standard, leap seconds are hard-coded into the Astropy source code. Therefore to include a new leap second the user must manually check for a new leap second and, if required, re-install and recompile Astropy! 

To make this process automated we have written a new
leap second management routine---{\tt barycorrypy.utc\textunderscore tdb()}, which we expose to users of the {\tt barycorrpy} package. This routine maintains its own leap second file in a user-defined directory, as well as a log-file which contains the date that leap second file was downloaded.

 Every time the routine is run, it uses the log-file to check the staleness of the leap second file, and alerts the user with a warning if the leap second file needs to be updated. Further if it should be updated and it cannot be (for instance, if there is no Internet access), the user may choose to ignore (or silence) the warning (though this would be advised against for applications that require 1 cm/s precision). In the absence of the ability to generate a new leap second file the routine defaults to the Astropy routine (albeit at the risk of missing leap second which would lead to loss of precision).  
 
 We finally note that {\tt barycorrpy} can be used as a standalone tool to calculate and include leap seconds, for instance when converting the UTC time scale to TDB, TT (Terrestrial Time) or Barycentric Julian Date TDB\footnote{The UTC to BJD$\_$TDB converter is in development and will be released in v0.2.} - \citet{Eastman}.

We would like to thank  Joe Ninan, Jason Eastman, Gudmundur Stefansson, Jiayin Dong, and Emily Lubar for their help in testing the code.

\begin{thebibliography}{}
\bibitem[Wright and Eastman (2014)]{JandJ} Wright, J. T. \& Eastman, J. D.\ 2014, \pasp, 126, 943 
\bibitem[Eastman, Siverd, and Gaudi (2010)]{Eastman} Eastman, J. \& Siverd R. \& Gaudi B.S. \ 2010, \pasp,  122, 935
\bibitem[Halverson et al.(2016)]{neid2} Halverson, S., Terrien, R., Mahadevan, S., et al.\ 2016, \procspie, 9908, 99086P 
\bibitem[Mahadevan et al.(2012)]{hpf} Mahadevan, S., Ramsey, L., Bender, C., et al.\ 2012, \procspie, 8446, 84461S 
\bibitem[Schwab (2016)]{neid} Schwab, C., Rakich, A., Gong, Q., et al.\ 2016, \procspie, 9908, 99087H 


\bibitem[Hobbs et al.(2006)]{Tempo} Hobbs, G.~B., Edwards, R.~T., \& Manchester, R.~N.\ 2006, \mnras, 369, 655 

\end{thebibliography}
\end{document}